\documentclass[a4paper, 11pt]{amsart}
\usepackage{fullpage}
\usepackage{amssymb,amsmath,latexsym,amsbsy,enumerate}
\usepackage{graphicx,subfigure}
\usepackage{epstopdf}
\usepackage{algorithm}
\usepackage{algorithmic}
\newenvironment{pf}{\begin{trivlist}
\item[\hspace{\labelsep}{\em\noindent Proof: }]
}{\hfill$\Box$\end{trivlist}}
\newtheorem{theo}{Theorem}
\newtheorem{lem}{Lemma}
\newtheorem{df}{Definition}

\begin{document}
\title{Level-k Phylogenetic Network can be Constructed from a Dense Triplet Set in Polynomial Time}
\author{Thu-Hien To \and Michel Habib}
\address{LIAFA, CNRS and University Paris Diderot - Paris 7}
\email{name@liafa.jussieu.fr}
\date{\today}
\maketitle
\begin{abstract}
Given a dense triplet set $\mathcal{T}$, there arise two interesting questions \cite{JS04}: Does there exists any phylogenetic network consistent with $\mathcal{T}$? And if so, can we find an effective algorithm to construct one?
For cases of networks of levels $k=0$ or $1$ or $2$, these questions were answered in \cite{ASSU81, JNS06, JS04, JS06, IKKS08} with effective polynomial algorithms. For higher levels $k$, partial answers were recently obtained in \cite{IK08} with an $O(|\mathcal{T}|^{k+1})$ time algorithm for simple networks. In this paper we give a complete answer to the general case, solving a problem of \cite{JS04}. The main idea is to use a special property of SN-sets in a level-k network. As a consequence, we can also find the level-k network with the minimum number of reticulations in polynomial time.
\end{abstract}

\section{Introduction}
The goal of phylogenetics is to reconstruct plausible evolutionary histories from biological data of currently living species. Normally, the standard model to describe the derivation is a tree whose each leaf is labeled by a specie and each node with descendants represents the most recent common ancestor of the descendants. But in reality, if we count to the hybridizations, recombinations and lateral gene transfer events, the model will be a network in which we allow the fact that a specie can have more than one parent. We call such a node a reticulation or a hybrid node. To study general phylogenetic networks, a way to classify them by level has been introduced in \cite{CJSS05}, basing on the number of reticulations in its biconnected components. A phylogenetic tree is considered as a level-0 phylogenetic network. This view gives us an approach to analyse networks thank to a beautiful structure so that we can decompose a network into several modules. In the other side, the most basic description of a phylogenetic evolution is a triplet which gives us the information on the relation of 3 species: which 2 species are closer than the last. Therefore, the considered problem is to construct a phylogenetic network consistent with a set of triplets. However, with an arbitrary triplet set, the problem is NP-hard with networks of levels higher than $0$ \cite{JNS06,IKKS08,IKM08}. But if we impose the density on the triplet set, which means that there is at least one triplet on each three species, then the triplet set has a better structure so that we can infer a level-1 \cite{JNS06, JS04, JS06}, or a level-2 \cite{IKKS08} network, if one exists, in polynomial time. The question firstly posed in \cite{JS04} is: Does the problem remain polynomial for level-k network with any $k$ fixed? We give an affirmative answer for this question here. As a consequence, we can also find the level-k network with the minimum number of reticulations, if one exists, in polynomial time.

\textbf{Related works:} Aho, Sagiv, Szymanski, and Ullman \cite{ASSU81} presented an $O(|\mathcal{T}|.n)$-time algorithm for determining whether a given set $\mathcal{T}$ of triplets on $n$ leaves is consistent with some rooted, distinctly leaf-labelled tree, i.e. a level-0 network, and if so, returning such a tree.
Later, there are improvements for this algorithm given in \cite{GJLO98, HKW99}.
But the problem becomes NP-hard for all other levels \cite{JNS06, IKKS08, IKM08}.
And the problem of finding a network consistent with the maximum number of triplets is also NP-hard for all levels \cite{JNS06,IKM08}.
The approximation problem which gives a factor on the number of triplets that we can construct a network consistent with, is also studied in \cite{BGHK08} for level-0, level-1, and level-2 networks.

Concerning the problems with dense triplet sets, there are following results.
For level-1, \cite{JNS06, JS04, JS06} give an $O(|\mathcal{T}|)$-time algorithm to construct a consistent network, and \cite{IK08} gives an $O(n^5)$-time algorithm to construct the consistent one with the minimum number of reticulations.
For level-2, \cite{IKKS08} gives an $O(|\mathcal{T}|^{\frac{8}{3}})$-time algorithm to construct a consistent network, and \cite{IK08} presents an $O(n^9)$-time algorithm to construct the consistent one with the minimum number of reticulations.
For level-k networks with any $k$ fixed, there is only a result for constructing all simple consistent networks with an $O(|\mathcal{T}|^{k+1})$-time algorithm \cite{IK08}.
The problem of finding a network consistent with the maximum number of triplets is also NP-hard for all levels in this case \cite{IKM08}.
However, it is still unknown if we can find the consistent networks with the minimum level in polynomial time.

There are also studies on the version of extremely dense triplet sets, that is when $\mathcal{T}$ is considered to contain all triplets of a network. In this case, an algorithm of $O(|\mathcal{T}|^{k+1})$ was given in \cite{IK08} for any level-k network. But even in this case, the problem of minimizing the level of consistent networks is still open.

\section{Preliminaries}
Let $\mathcal{L}$ be a set of $n$ species. A \textit{phylogenetic network} $N$ on $\mathcal{L}$ is a connected, directed, acyclic graph which has:

- a unique vertex of indegree $0$ and outdegree $2$ (root).

- vertices of indegree $1$ and outdegree $2$ (speciation vertices).

- vertices of indegree $2$ and outdegree $1$ (reticulation vertices, or hybrid vertices).

- $n$ vertices labelled distinctly by $\mathcal{L}$ of indegree $1$ and outdegree $0$ (leaves). So $\mathcal{L}$ is also called the leaf set.

We denote $u\leadsto v$ if there is a path in $N$ from $u$ to $v$ ($u$ and $v$ may be the same vertex).

A graph is \textit{biconnected} if it contains no vertex whose removal disconnects the graph.
A biconnected component of a graph is a maximal biconnected subgraph.
Let $\mathcal{U}(N)$ be the underlying undirected graph of $N$, obtained by replacing each directed edge of $N$ by an undirected edge. We consider the decomposition into biconnected components of $\mathcal{U}(N)$. As any two biconnected components of $\mathcal{U}(N)$ are vertex-disjoint, $\mathcal{U}(N)$ consists of a finite number of vertex-disjoint biconnected components. Each remaining edge connects two biconnected components. In $N$ such edge correponds to an arc whose removal disconnects $N$. So we call it a \textit{cut-arc}. A cut-arc $a=(u,v)$ is \textit{highest} if there is no cut-arc $a'=(u',v')$ such that $v'\leadsto u$.

A network $N$ is called of \textit{level-$k$} if every biconnected component of $\mathcal{U}(N)$ contains at most $k$ hybrid vertices.

A \textit{triplet} $x|yz$ is a rooted binary tree on the leaves $x$, $y$ and $z$ such that $x$ and the parent of $y$ and $z$ are children of the root. A set $\mathcal{T}$ of triplets is \textit{dense} if for any set $\{x,y,z\}\subseteq \mathcal{L}$, at least one triplet on these three leaves belongs to $\mathcal{T}$.

A triplet $x|yz$ is \textit{consistent} with a network $N$ if $N$ contains two vertices $u \not= v$ and pairwise internally vertex-disjoint paths $u\leadsto x$, $u\leadsto v$, $v\leadsto y$, and $v\leadsto z$.

A phylogenetic network is \textit{simple} if it has only one non leaf biconnected component which is the one containing the root, and every its cut-arc connects a vertex of this biconnected component to a leaf.

Let $\mathcal{P}$ be a partition of the leaf set $\mathcal{L}$: $\mathcal{P}=\{P_1, \dots, P_q\}$. We denote $\mathcal{T} \nabla \mathcal{P}$ the induced set of triplets $P_iP_j|P_k$ such that there exist $x \in P_i, y \in P_j, z \in P_k$ with $xy|z \in \mathcal{T}$ and $i, j$ and $k$ are distinct.

\begin{figure}[ht]
\begin{center}
\includegraphics[scale=0.7]{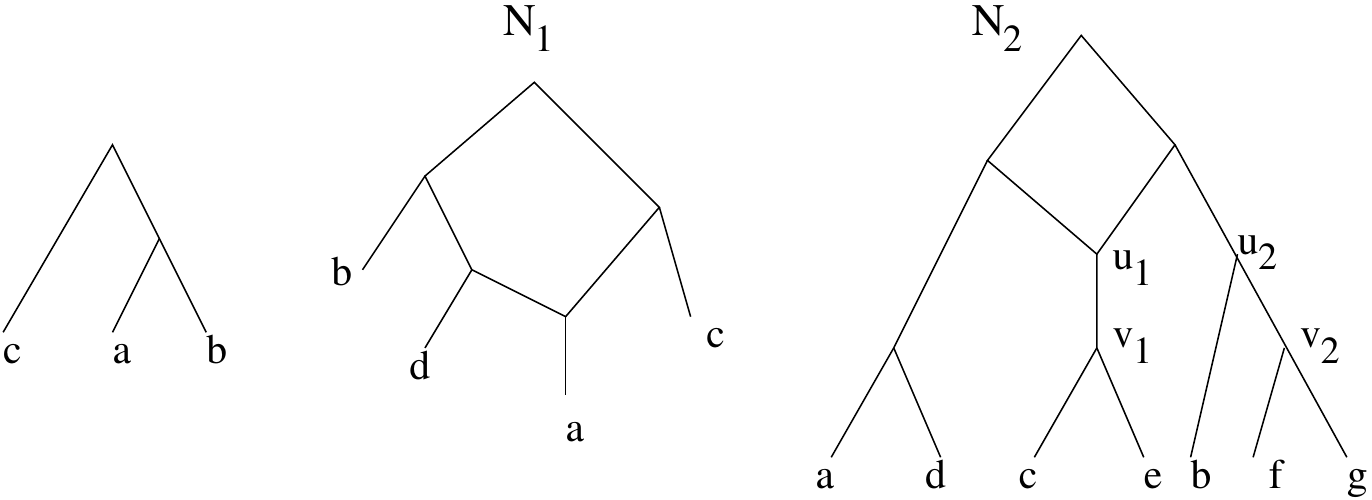}
\caption{The triplet $c|ab$ is consistent with $N_1$, but not with $N_2$.
$N_1$ is a simple level-1 network, $N_2$ is also a level-1 network but not simple.
In $N_2$, $(u_1,v_1)$ is a highest cut-arc, $(u_2, v_2)$ is also a cut-arc but not highest.
Note that, as with all figures in this article, all arcs are directed downwards, away from the root.
}
\label{preliminaries}
\end{center}
\end{figure}

\section{Construction a Level-k Phylogenetic Network from a Dense Triplet Set}
In this section we show that with any $k$ fixed, it is possible to construct in polynomial time a level-k phylogenetic network from a dense triplet set, if such a network exists. Let us start with some properties of level-k networks.

\begin{figure}[ht]
  \label{construction}
  \subfigure[Decomposition of a network $N$: the biconnected component $C$ is in bold, each sub-network $N_j$ is framed by a dotted bold rectangle, each highest cut-arc connects from the biconnected component to a sub-network. \label{decomposition}]{\includegraphics[scale=0.75]{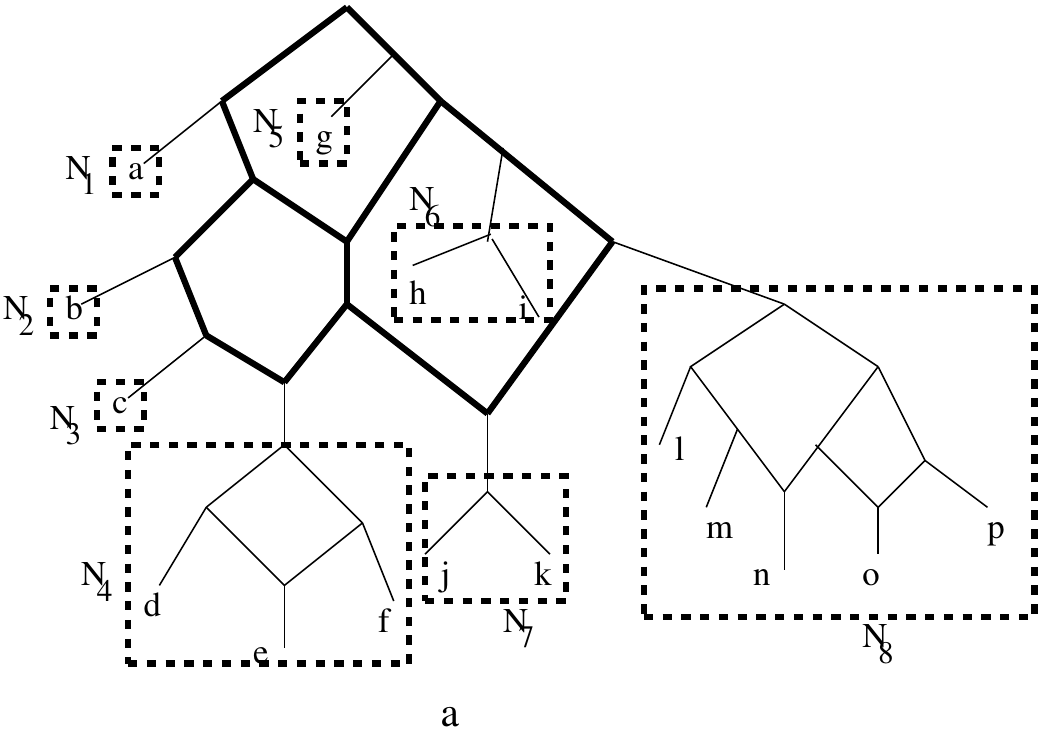}}\; \; \; \; \;
  \subfigure[The corresponding simple network $N_s$ of $N$. Each leaf $P_j$ represents the leaf set of the sub-network $N_j$. \label{simple_network}]{\includegraphics[scale=0.75]{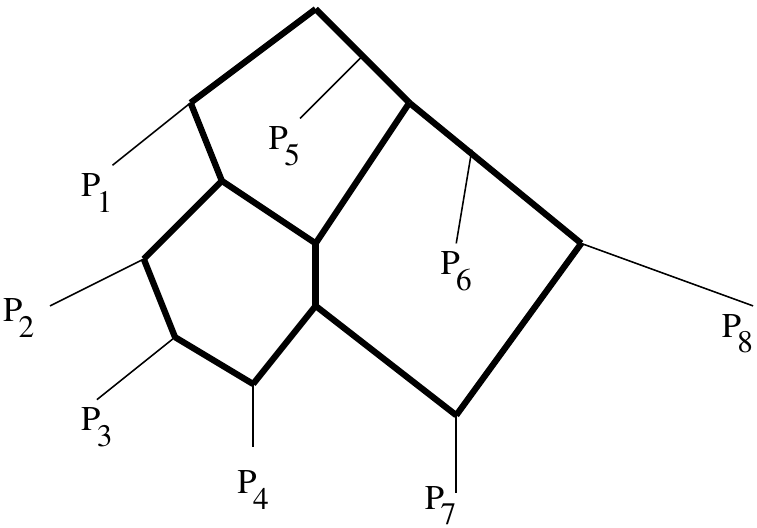}}
  \caption{Construction a network by a recursive algorithm}
\end{figure}

Let $N$ be a level-k network. Then:

i) A cut-arc connects two vertex-disjoint sub-networks of $N$ and each one is also a level-k network.

ii) We can decompose $N$ into a finite number of modules as follows (see figure \ref{decomposition}): One of the modules is a biconnected component $C$ which contains the root. The other modules are level-k sub-networks $N_1,\dots ,N_m$, which are pairwise vertex-disjoint. Moreover, for any $j=1,\dots ,m$, there is a unique arc, called a highest cut-arc, connecting from $C$ to $N_j$.

iii) For any $j=1,\dots ,m$, let $P_j$ be the leaf set of the sub-network $N_j$. So $\mathcal{P} = \{P_1, \dots ,P_m\}$ is a partition of $\mathcal{L}$. If we replace each $N_i$ by a representing leaf, also called $P_i$, we obtaine a simple network $N_s$ (see figure \ref{simple_network}). Asumming that for any $j$, $N_j$ is consistent with $\mathcal{T}|P_j$. So, $N$ is consistent with $\mathcal{T}$ if and only if $N_s$ is consistent with $\mathcal{T} \nabla \mathcal{P}$.

Using these properties, we can have the following recursive algorithm. Firstly, look for the possible decompositions: how the leaf set is partitioned below the highest cut-arcs and what is the consistent simple network whose each leaf represents a part of the partition. Then, recursively construct a consistent sub-network on each part of the partition.
We know that it is possible to construct all simple networks consistent with a dense triplet set $\mathcal{T}$ in $O(|\mathcal{T}|^{k+1})$ time \cite{IK08}.
So it remains to know the possible partitions of the leaf set $\mathcal{L}$ below the highest cut-arcs. We will show in the remaining of this section that the number of the possible partitions is bound by a polynomial function of $n$.
This fact allows us to construct a level-k network, and then a level-k network which minimizes the number of reticulations in polynomial time.

The question is answered by exploring the leaf sets hung below cut-arcs. Remark that if $A$ is a leaf set hung below a cut-arc, then for any $z \in \mathcal{L} \backslash A$, $x, y\in A$, the only triplet on $\{x,y,z\}$ that can be consistent with the network is $z|xy$. Basing on this property, we define a family of leaf sets, called \textit{CA-sets}, for CutArc-sets, as follows.

\begin{df}
\label{CA-set}
Let $A \subseteq \mathcal{L}$. We say that $A$ is a CA-set if either it is a singleton or the whole $\mathcal{L}$, or if it satisfies the following property: For any $z \in \mathcal{L} \backslash A$, $x, y\in A$, the only triplet on $\{x,y,z\}$ in $\mathcal{T}$, if there is any, is $z|xy$.
\end{df}

As remarked, a leaf set hung below a cut-arc is a CA-set, but the converse sens is not always true. Let us recall  that \cite{JS04} presented  a variation  of these CA-sets, namely the notion of SN-set. A \textit{SN-set} is defined on a leaf set. Let $A$ be a subset of $\mathcal{L}$, the SN-set of $A$,  denoted $SN(A)$,  is the set recursively defined as $SN(A \cup \{z\})$ if there exists some $z \in \mathcal{L} \backslash A$ and $x, y \in A$ such that $x|yz \in \mathcal{T}$, and as $A$ otherwise. We will show in the following lemma that the two definitions identify the same family of leaf sets.

\begin{lem} Equivalence of the two definitions.
\label{equi-def}

(i) For any $A \subseteq \mathcal{L}$, $SN(A)$ is a CA-set.

(ii) For any CA-set $A$, there exists $B$, a subset of $\mathcal{L}$, such that $SN(B) = A$.
\end{lem}

\begin{pf}
All claims are obviously true with singleton sets. So we consider only the non singleton sets in the next.

(i) For any non singleton set $A \subseteq \mathcal{L}$, $\forall z \in \mathcal{L} \backslash SN(A)$, $\forall x, y \in SN(A)$, neither $x|yz$ nor $y|xz$ is in $\mathcal{T}$ because if one of them is, following the definition of SN-set, $SN(A)$ will be $SN(A \cup \{z\})$, and will contains $z$. So, the only triplet on $\{x,y,z\}$ in $\mathcal{T}$, if there is any, is $z|xy$. Or, $SN(A)$ is a CA-set, according to the definition \ref{CA-set}.

(ii) For any CA-set $A$, there can exist several $B$ such that $SN(B) = A$. We take, for exemple, $B$ equals to $A$. We have to show that $SN(A) = A$. Indeed, as $A$ is a CA-set, there doesn't exist any $z \in \mathcal{L} \backslash A$ and $x, y \in A$ such that $x|yz \in \mathcal{T}$. It means that $SN(A)$ is exactly $A$, accoding to the recursive definition of SN-set.
\end{pf}

Therefore, the family of SN-sets is exactly the family of CA-sets and  we will stick to  the notation of SN-set for any CA-set determined by the definition \ref{CA-set}.

It was proved in \cite{JS04} that if $\mathcal{T}$ is dense, then the collection of the SN-sets is a laminar family \cite{S03}.
It means that two SN-sets are either disjointed or included one in another, and the family is tree structured under inclusion.
So all SN-sets are representable by a tree, called SN-tree.
Each node of SN-tree corresponds to a SN-set.
The root corresponds to $\mathcal{L}$, and the leaves correspond to the singletons.
The SN-tree can be calculated in $O(n^3)$ time \cite{JNS06}.

Let $A, a$ be two SN-sets.
We say that $a$ is a \textit{child} of $A$ if in the SN-tree, the node which represents $a$ is a child of the node which represents $A$.

\begin{figure}[ht]
  \label{def-partition}
  \subfigure[The SN-tree of $\mathcal{T}$ \label{SN-tree}]{\includegraphics[scale=0.5]{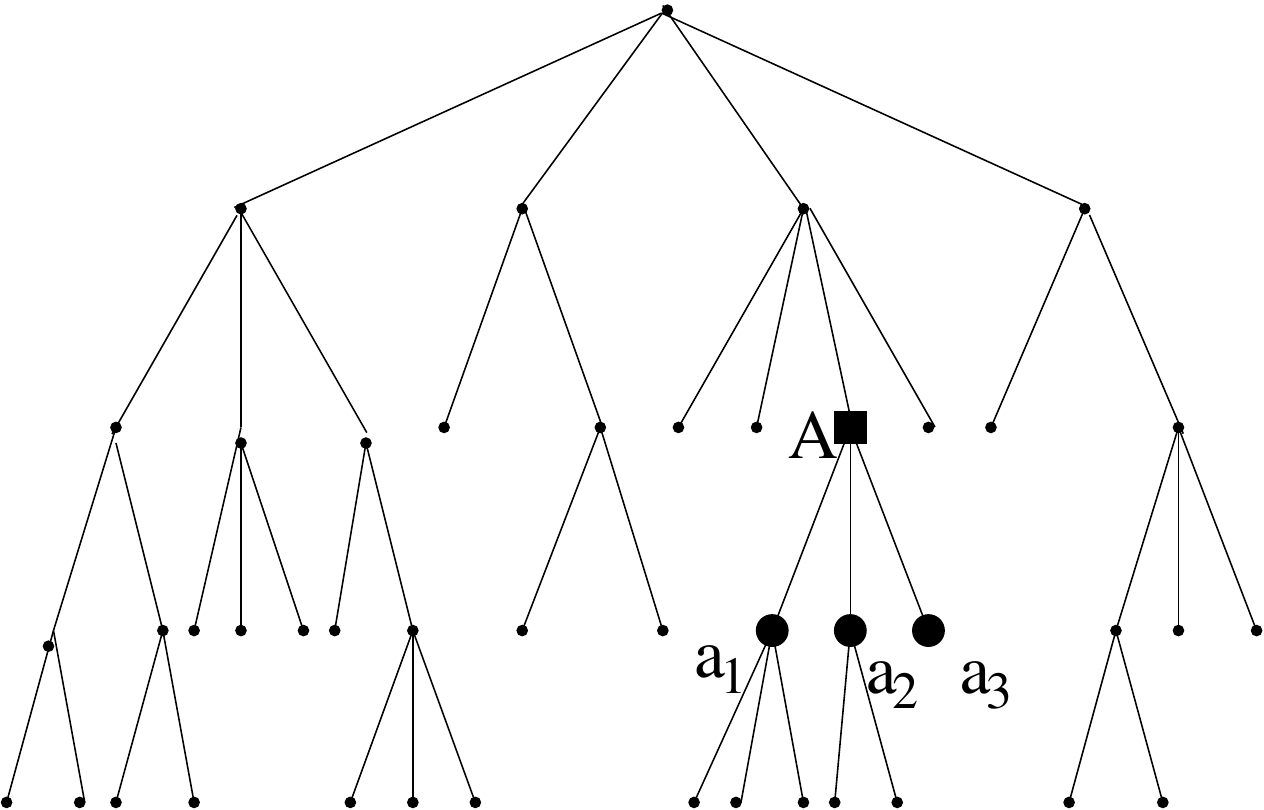}}\; \; \; \;
  \subfigure[a network $N$ consistent with $\mathcal{T}$ \label{N}]{\includegraphics[scale=0.5]{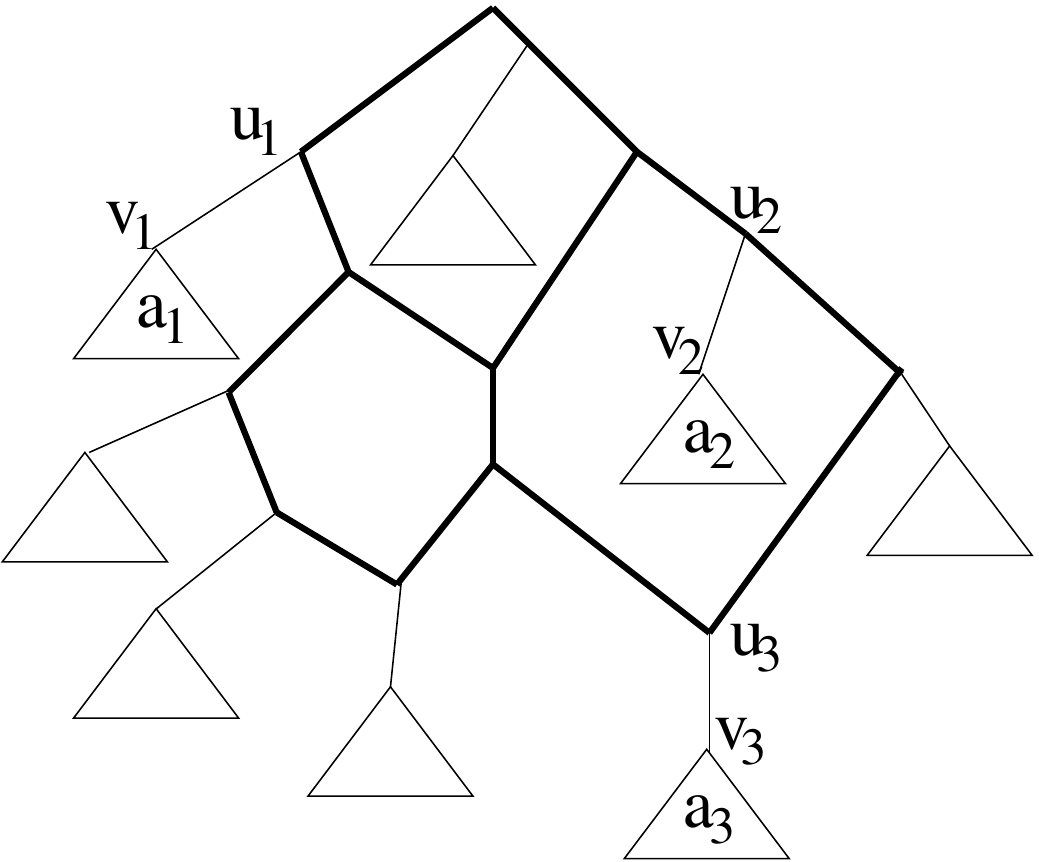}}
  \caption{The SN-set $A$ is splitted in $N$. Each leaf set $a_i$, child of $A$, is hung below a highest cut-arc $(u_i,v_i)$ of $N$.}
\end{figure}

\begin{df}
Let $N$ be a network consistent with $\mathcal{T}$, and $A$ be a SN-set.
We say that $A$ is \textbf{splitted} in $N$ if each of its children is hung below a highest cut-arc of $N$ (see figure \ref{def-partition}).
\end{df}

Following the definition, if $A$ is splitted, its children are not.

\begin{figure}[ht]
\begin{center}
\includegraphics[scale=0.5]{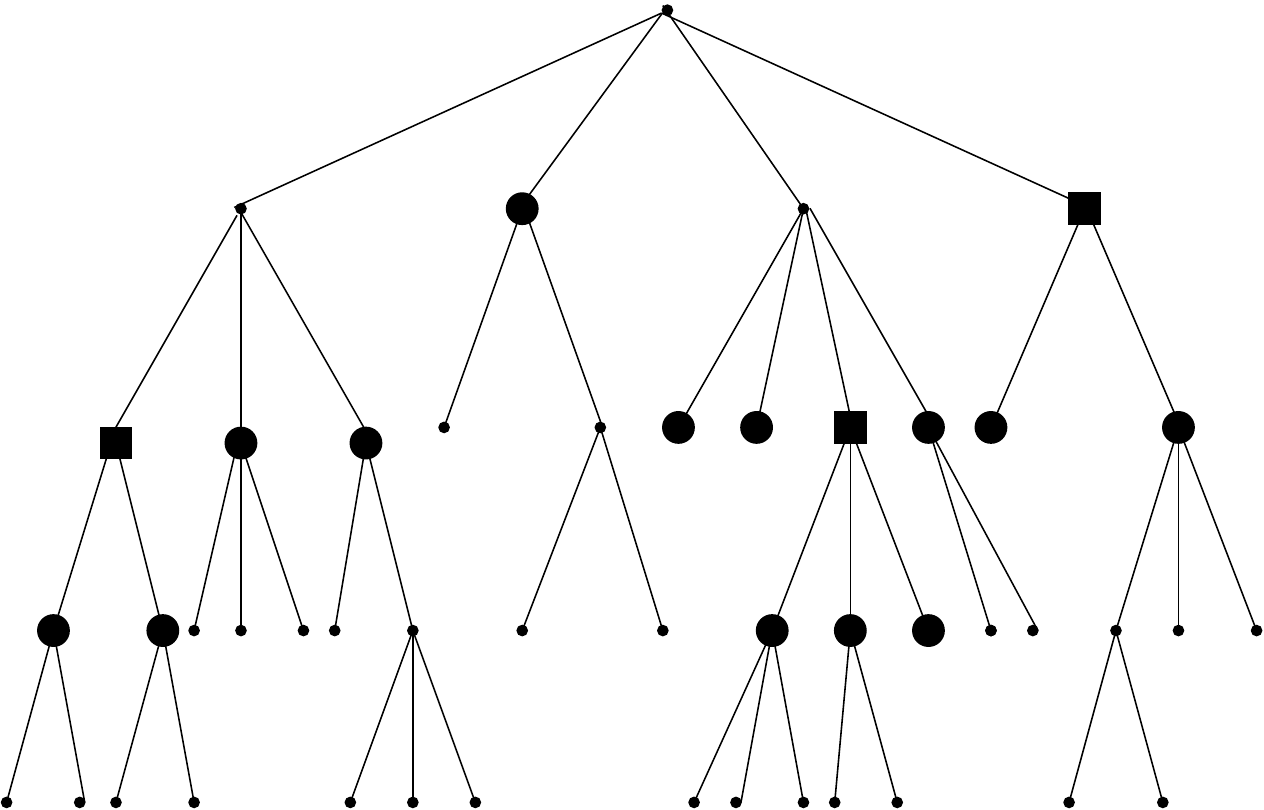}
\caption{The black square nodes represent the splitted SN-sets. The black round nodes represent the SN-sets hung below the highest cut-arcs.}
\label{SN-arbre}
\end{center}
\end{figure}

Let us start with a simple  remark: \emph{The knowledge of all splitted SN-sets is enough to capture all the SN-sets hung below the highest cut-arcs, or to capture the partition of the leaf set.}
Really, let see the example of a SN-tree in the figure \ref{SN-arbre}, the black square nodes represent the splitted SN-sets, we have three.
The children of these three sets, with the maximal SN-sets that do not contain any of these three sets are the SN-sets hung below the highest cut-arcs.
In the figure, these sets are marked by the black round nodes, they create a partition of $\mathcal{L}$.

Let $\mathcal{T}$ be a dense triplet set. $N$ is a level-k network consistent with all triplets of $\mathcal{T}$. $N_S$ is the simple network of $N$. $H$ is the set of the hybrid vertices of $N_S$, so $|H| \leq k$. $\mathcal{H}$ is the set of all subsets of $H$.
$\mathcal{A}$ is the set of all splitted SN-sets in $N$ (In the figure \ref{SN-arbre}, $\mathcal{A}$ is the set of the SN-sets corresponding to the square nodes ). We define a function $f$ from $\mathcal{A}$ to $\mathcal{H}$ as follows.

\begin{figure}[ht]
\begin{center}
\includegraphics[scale=0.5]{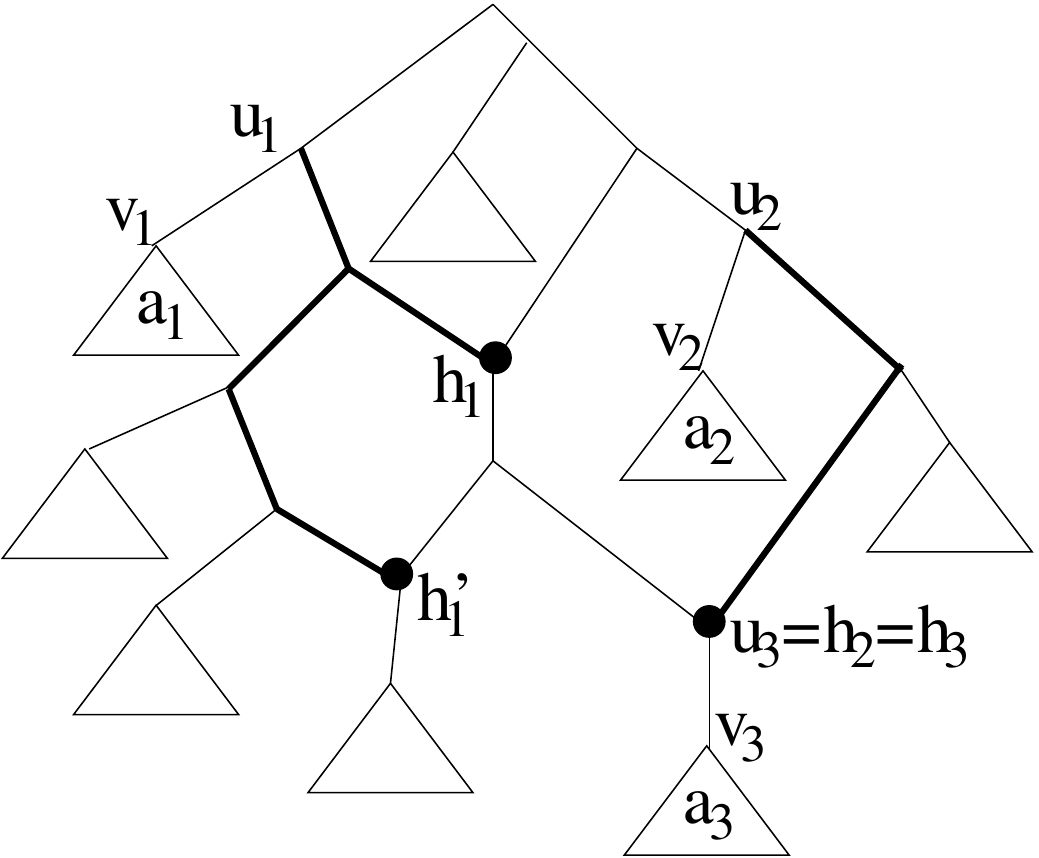}
\caption{$f(A) = \{h_1,h_1',u_{3}\}$ where $A$ is the splitted SN-set which has three children $a_1, a_2, a_3$. The paths, which are in bold, $u_1 \leadsto h_1$, $u_1 \leadsto h_1'$, and $u_2 \leadsto u_3$ don't contain any internal hybrid vertex.} 
\label{function_f}
\end{center}
\end{figure}

\begin{df}
Given $A \in \mathcal{A}$ and $a_1, \dots a_m$ the children of $A$. In $N$, each $a_i$ is hung below a highest cut-arc $(u_i,v_i)$. We define:

$f(A)=\{h \in H |\exists i$ so that $u_i \leadsto h$ and the path from $u_i$ to $h$ does not contain any internal hybrid vertex (if $u_i$ is a hybrid vertex, then $h = u_i$)$\}$ (see figure \ref{function_f}).

\end{df}

\begin{lem}
\label{lem1}
The function $f$ has the following properties:

(i) $\forall A \in \mathcal{A}, f(A) \neq \emptyset$.

(ii) $\forall h \in H$, there are at most three pairwise disjointed sets of $\mathcal{A}$ so that their image by $f$ contains $h$.
\end{lem}

\begin{pf}
(i) For any $A \in \mathcal{A}$, we prove that $\forall u_i, \exists h \in H$ such that $u_i \leadsto h$ and the path from $u_i$ to $h$ does not contain any internal hybrid vertex. This fact implies $f(A) \neq \emptyset$.

Indeed, if $u_i$ is a hybrid vertex, then we have $h = u_i$.

If $u_i$ is not a hybrid vertex, then there are two arcs starting from $u_i$: one is $(u_i,v_i)$ and let the second be $(u_i,v_i')$. Assuming that there is no hybrid vertex of $N_S$ that is reachable from $u_i$, so both $(u_i,v_i)$ and $(u_i,v_i')$ are cut-arcs. We infer that the arc coming to $u_i$ is also a cut-arc, so $(u_i,v_i)$ is not a highest cut-arc, a contradiction.

(ii)
Assuming that there are four pairwise disjointed sets $A_1, A_2, A_3, A_4 \in \mathcal{A}$ so that $\exists h \in f(A_1) \cap f(A_2) \cap f(A_3) \cap f(A_4)$.  According to the definition of $f$, $\forall i \in \{1,2,3,4\}$, there is at least a child $a_i$ of $A_i$ so that no internal vertex of the path from $u_{i}$ to $h$ is hybrid.

Firstly, among the four $u_i's$, there is at most one that is equal to $h$. It means that there are at least three $u_i's$ that are strictly above $h$. As $h$ has only two parents, and the path from $u_i$ to $h$ does not contain any internal hybrid vertex, so there exist $i1,i2 \in \{1,2,3,4\}$ so that $u_{i2}$ is placed on the path from $u_{i1}$ to $h$.

The following proof is illustrated by the figure \ref{lemme}.

\begin{figure}[ht]
\begin{center}
\includegraphics[scale=0.8]{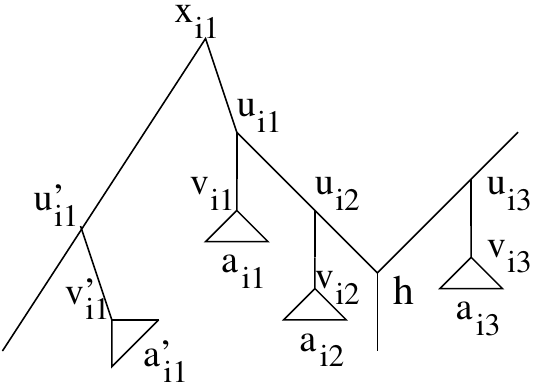}
\caption{The triplets $a_{i2}|a_{i1}a_{i1}'$ can not be consistent with the network.}
\label{lemme}
\end{center}
\end{figure}
For convenience, we call the triplets $a_{i}|a_{j}a_{k}$ for the set of all triplets $x|yz$ where $x \in a_i, y \in a_j$ and $z \in a_k$.

Let $a_{i1}'$ be another child of $A_{i1}$. As $A_{i1}$ is a SN-set, and $a_{i2}$ is not included in $A_{i1}$ (because $A_{i1}$ and $A_{i2}$ are disjointed), so according to the definition, the triplets $a_{i2}|a_{i1}a_{i1}'$ have to be contained in $\mathcal{T}$. Let $x_{i1}$ be any common ancestor of $u_{i1}$ and $u_{i1}'$, then $x_{i1}$ is above $u_{i1}$. We remark that all paths starting from a vertex above $u_{i1}$ that come to $a_{i2}$ have to pass by $u_{i1}$ because there is no hybrid vertex on the path from  $u_{i1}$ to $u_{i2}$. Besides, all paths coming to $a_{i1}$ have to pass by $u_{i1}$ too. Then the triplets $a_{i2}|a_{i1}a_{i1}'$ can not be consistent with the network. So $a_{i2}|a_{i1}a_{i1}'$ is not contained in $\mathcal{T}$, contradiction.
\end{pf}

Therefore, we have the following lemma that allows us to bound the number of splitted SN-sets in a level-k network.

\begin{lem} (Fundamental)
\label{lem2}

Let $\mathcal{T}$ be a dense triplet set which is consistent with a level-k network $N$, $\mathcal{A}$ be the collection of splitted SN-sets in $N$, then $|\mathcal{A}| \leq 3k$.
\end{lem}

\begin{pf}
Firstly, we observe that all elements of $\mathcal{A}$ are pairwise disjointed. Actually, for some two SN-sets, they are either disjointed or included one in another. But according to the definition, if $A$ is splitted, its children are not. It means that if $A$ is in $\mathcal{A}$, its subsets are not. Then, the two sets of $\mathcal{A}$ can not be included one in another. They are disjointed.

Let $H' \subseteq H$ be the union of the images of all elements of $\mathcal{A}$ by $f$, so $|H'| \leq k$. 
Let $A \in \mathcal{A}$, and $h \in H'$, we say that $A$ corresponds to $h$, and $h$ corresponds to $A$ if $f(A)$ contains $h$. 
We infer from the lemma \ref{lem1} that each element of $\mathcal{A}$ corresponds to at least one element of $H'$, and each element of $H'$ corresponds to at most three elements of $\mathcal{A}$. So $|\mathcal{A}| \leq 3|H'| \leq 3k$.
\end{pf}

Moreover, if we are interested only in finding a certain consistent network, we can have a better bound with the following lemma. The idea is to modify an arbitrary consistent network into another one which has a particular property but still consistent with $\mathcal{T}$ and don't make the level increase. But the class of  modified networks will not be assured to contain the one with the minimum number of reticulations.

\begin{lem}
\label{lem3}
Let $\mathcal{T}$ be a dense triplet set, if $\mathcal{T}$ is consistent with a level-k network $N$ whose 
the corresponding simple network has level greater than $1$, then there exists a level-k network $N'$ consistent with $\mathcal{T}$ such that: For any splitted SN-set $A$ of $N'$, $|f(A)| \geq 2$.
\end{lem}

\begin{pf}
Assuming that there exists a SN-splitted $A$ of $N$ such that $|f(A)| = 1$. Let $f(A)=\{h\}$, and $G_A$ be the sub-network of $N$ on $A$.
In $N_S$, there are $2$ paths leading to $h$. So there are $2$ cases that can happen.

In the first case (figure \ref{lem3a}), $u_i$ are all placed on one path leading to $h$, for example on the left one. Let $u_1$ be the highest and $u_f$ be the lowest vertex on all $u_i$. There are two possible positions for $u_f$: either it is right above $h$, i.e $(u_f,h)$ is an arc, or it is equal to $h$. The network $N'$ is obtained from $N$ by the following modifications: deleting all children $a_i$ of $A$ and all concerning arcs and vertices; at the position of $u_f$, add a new arc which connect to the sub-network $G_A$ at $u_1$ (figure \ref{lem3b}).

In the second case (figure \ref{lem3c}), $u_i$ are placed on the two paths leading to $h$. We can easily remark that the leaf set hung below $h$ has to be also a child of $A$.
The network $N'$ is obtained from $N$ by the following modifications: deleting all children $a_i$ of $A$ and all concerning arcs and vertices;  let $G_A'$ be the network obtained from $G_A$ by sticking the top of the two branches of $G_A$ into one vertex $u$. At the position of $h$, we add a new arc which connect to $G_A'$ at $u$ (figure \ref{lem3d}).

\begin{figure}[ht]
  \subfigure[\label{lem3a}]{\includegraphics[scale=0.85]{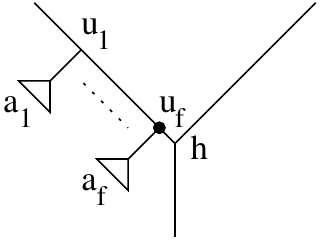}}\; \; \; \;
  \subfigure[The modified network of (a) \label{lem3b}]{\includegraphics[scale=0.85]{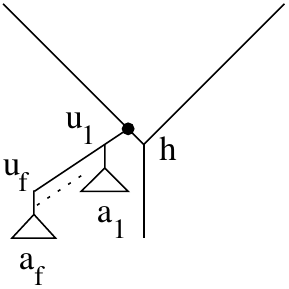}}\; \; \; \;
  \subfigure[\label{lem3c}]{\includegraphics[scale=0.85]{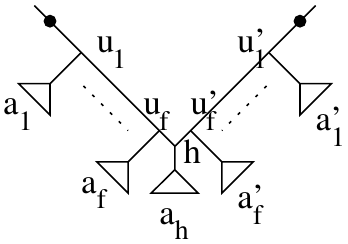}}\; \; \; \;
  \subfigure[The modified network of (c) \label{lem3d}]{\includegraphics[scale=0.85]{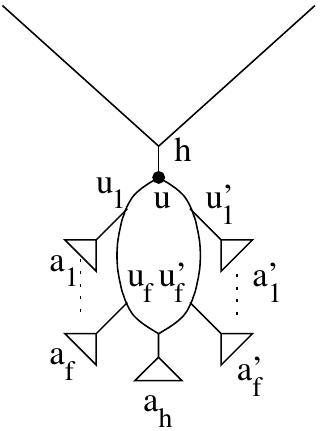}}\; \; \; \;
  \caption{The modified networks are level-k, still consistent with $\mathcal{T}$, and have all sons of $A$ hung below a highest cut-arc.}
\end{figure}

In the two cases, we can verify that the modifications don't increase the level of the network,  the new network is still consistent with all triplets of $\mathcal{T}$, and $A$ is not anymore a splitted SN-set of the new network because it is now hung below a highest cut-arc. The fact that the corresponding simple network of $N$ has level greater than $1$ assures that the new network doesn't contain any two parallel arcs with the sames extremities.

By modifying the network for any splitted SN-set of $N$ whose image by $f$ contains only one element, we obtain finally a network in which there is not anymore such splitted SN-set. In addition, the lemma \ref{lem1} says that the image by $f$ of any splitted SN-sets is not empty. Then we have a new network in which $|f(A)| \geq 2$ for any splitted SN-set $A$.
\end{pf}

\begin{lem}
\label{lem4}
Let $\mathcal{T}$ be a dense triplet set, if $\mathcal{T}$ is consistent with a level-k network $N$, then there exists a level-k network $N'$ consistent with $\mathcal{T}$ which satisfies: let $\mathcal{A}$ be the collection of splitted SN-sets in $N'$, then $|\mathcal{A}| \leq \lfloor\frac{3k}{2}\rfloor$.
\end{lem}

\begin{pf}
If the simple network of $N$ is of level-1, we choose $N' = N$. It can be inferred from \cite{JNS06, JS04} that each SN-set hung below a highest cut-arc is a son of the SN-set $\mathcal{L}$. It means that there is only one splitted SN-set $\mathcal{L}$. So $|\mathcal{A}| = 1 \leq \lfloor\frac{3k}{2}\rfloor$ is obviously true in this case.

Otherwise, according to the lemma \ref{lem3}, there exists a level-k network $N'$ consistent with $\mathcal{T}$ and satisfies: if $\mathcal{A}$ is the collection of splitted SN-sets of $N'$, then $\forall A \in \mathcal{A}, |f(A)| \geq 2$.
Let $H' \subseteq H$ be the union of the images of all of the elements of $\mathcal{A}$ by $f$, so $|H'| \leq k$. 
We say that $A$ corresponds to $h$, and $h$ corresponds to $A$ if $f(A)$ contains $h$.
So each element of $\mathcal{A}$ corresponds to at least two elements of $H'$ (lemma \ref{lem3}), and each element of $H'$ corresponds to at most three elements of $\mathcal{A}$ (lemma \ref{lem1}). 
Then $|\mathcal{A}| \leq \lfloor\frac{3}{2}|H'|\rfloor \leq \lfloor\frac{3k}{2}\rfloor$.
\end{pf}

\begin{theo}
Given a dense triplet set $\mathcal{T}$, and $k \geq 0$, it is possible to construct a level-k network consistent with $\mathcal{T}$, if one exists, in time $O(|\mathcal{T}|^{k+1}n^{\lfloor\frac{3k}{2}\rfloor+1})$.
\end{theo}

\begin{algorithm}
\caption{Level-k network}
\label{algo}
\begin{algorithmic}
\REQUIRE A dense triplet set $\mathcal{T}$.
\ENSURE A level-k network consistent with $\mathcal{T}$, if one exists; otherwise, $null$.
\STATE Calculate the SN-tree $R$ of $\mathcal{T}$.
\STATE  For every leaf $u$ of $R$, define $N_u$ the network contains only one leaf $u$.
\FOR {each internal node $a$ of $R$, in bottom-up order}
	\STATE Let $R[a]$ be the subtree of $R$ rooted at $a$.
	\STATE Let $n(a)=\{a_1, a_2, \dots , a_q\}$ be all of the nodes of $R[a]$
	\STATE For any $j = \{1, \dots ,q\}$, $n_j$ be the leaf set of $R[a_j]$.
	\STATE $i=1$; $found = false$;
	\WHILE {($i \leq \frac{3k}{2}$) and ($i \leq q$) and $!(found)$}
		\FOR {each combination $\mathcal{A}$ of $i$ disjointed sets $n_j$}
			\STATE Calculate the partition $\mathcal{P}$ of the leaf set of $R[a]$ from $\mathcal{A}$.
			\STATE Calculate $\mathcal{T'}$ from $\mathcal{T}|a$.
			\STATE Calculate $\mathcal{T'} \nabla \mathcal{P}$.
			\STATE Look for a certain simple network consistent with $\mathcal{T'} \nabla P$. If one exists, then denote $Ns_a$ this network; replace each leaf $f$ of $Ns_a$ by the network $N_f$ already found to obtain the network $N_a$; $found=true$; break.
		\ENDFOR	
			\IF {$found=true$}
				\STATE break;
			\ENDIF
			\STATE $i++$;
	\ENDWHILE
	\IF {$found=false$}
		\STATE \textbf{return} \textit{null};
	\ENDIF
\ENDFOR
\STATE \textbf{return} $N_r$ where $r$ is the root of $R$.
\end{algorithmic}
\end{algorithm}
\begin{pf}
Each element of $\mathcal{A}$ has children, so is not a singleton. The number of non singleton SN-sets is $O(n)$. So, from the lemma \ref{lem4}, we have $O(n^{\lfloor\frac{3k}{2}\rfloor})$ possible possibilities of $\mathcal{A}$ by choosing at most $\lfloor\frac{3k}{2}\rfloor$ disjointed SN-sets from all non singleton SN-sets. For each choice of $\mathcal{A}$, we look for the corresponding partition $\mathcal{P}$ of $\mathcal{L}$. Each part of the partition is either a child of an element of $\mathcal{A}$, or a maximal SN-sets that does not contain any element of $\mathcal{A}$.

Next, we have to find a simple network consistent with $\mathcal{T} \nabla \mathcal{P}$. The theorem $3$ in \cite{IK08} says that it is possible to construct all simple level-k networks consistent with a dense triplet set $\mathcal{T}$ in $O(|\mathcal{T}|^{k+1})$ time. So, it takes totally $O(|\mathcal{T}|^{k+1}n^{\lfloor\frac{3k}{2}\rfloor})$ time to find all the possible decompositions.
As a recursive algorithm, we suppose that for each part $P_i$ of the partition $\mathcal{P}$, we already knew a network consistent with $\mathcal{T}|P_i$, if there is any. Then, the wanted network will be obtained by replacing each leaf which represents the part $P_i$ of the simple network by the corresponding network on $P_i$.

The algorithm, which is described in the algorithm \ref{algo}, constructs on each non singleton SN-set, in small-big order, a consistent network, if there is any. If there isn't any such network, we can conclude immediately that there isn't any network consistent with $\mathcal{T}$, and the algorithm returns $null$. Indeed, if there exists a network $N$ consistent with $\mathcal{T}$, then for any SN-set $A$ there is always a network consistent with $\mathcal{T}|A$ which is the sub-network of $N$ on $A$. The last obtained network, on the largest SN-set which is $\mathcal{L}$, is the wanted one. As there are $O(n)$ non singleton SN-sets, the complexity will be multiplied by $n$. The construction of SN-tree takes $O(n^3)$, all other operations take a negligible time compared with the others. So, the total complexity is $O(|\mathcal{T}|^{k+1}n^{\lfloor\frac{3k}{2}\rfloor+1})$.
\end{pf}

As a consequence, and with a recursive property of the network with minimum number of hybrid vertices, the problem of finding the consistent network which minimizes the number of hybrid vertices is also solved in polynomial time.

\begin{theo}
Given a dense triplet set $\mathcal{T}$, and $k \geq 0$, it is possible to construct the level-k network consistent with $\mathcal{T}$ which minimizes the number of hybrid vertices, if one exists, in time $O(|\mathcal{T}|^{k+1}n^{3k+1})$.
\end{theo}

\begin{algorithm}
\caption{Level-k network with the minimum number of hybrid vertices}
\label{algo_min}
\begin{algorithmic}
\REQUIRE A dense triplet set $\mathcal{T}$.
\ENSURE A level-k network consistent with $\mathcal{T}$ that minimizes the number of hybrid vertices, if one exists; $null$ otherwise.
\STATE Calculate the SN-tree $R$ of $\mathcal{T}$.
\STATE For every leaf $u$ of $R$, define $N_{u_{min}}$ the network contains only one leaf $u$.
\FOR {each internal node $a$ of $R$, in bottom-up order}
	\STATE Let $R[a]$ be the subtree of $R$ rooted at $a$.
	\STATE Let $n(a)=\{a_1, a_2, \dots , a_q\}$ be all of the nodes of $R[a]$.
	\STATE $N_{a_{min}} = null$;$min = n$;
	\STATE $i=1$;
	\WHILE {($i \leq 3k$) and ($i \leq q$)}
		\FOR {each combination $\mathcal{A}$ of $i$ disjointed elements of $n(a)$}
			\STATE Calculate the partition $\mathcal{P}$ from $\mathcal{A}$.
			\STATE Calculate $\mathcal{T'}$ the triplet set on the leaves of $R[a]$.
			\STATE Calculate $\mathcal{T'} \nabla \mathcal{P}$.
			\FOR {each simple network $Ns_a$ consistent with $\mathcal{T'} \nabla P$}				
				\STATE Construct $N_a$ by replacing each leaf $f$ of $Ns_a$ by the network $N_{f_{min}}$ already found.
				\STATE $m =$ the number of hybrid vertices of $N_a$.
				\IF {$m < min$}
					\STATE $min = m$;$N_{a_{min}} = N_a$;
				\ENDIF
			\ENDFOR
		\ENDFOR	
			\STATE $i++$;
	\ENDWHILE
	\IF {$N_{a_{min}}=null$}
		\STATE \textbf{return} \textit{null};
	\ENDIF
\ENDFOR
\STATE \textbf{return} $N_{r_{min}}$ where $r$ is the root of $R$.
\end{algorithmic}
\end{algorithm}
\begin{pf}
Let $N$ be a level-k network consistent with $\mathcal{T}$, $\mathcal{P} = (P_1, \dots , P_m)$ be the partition of the leaf set hung below the highest cut-arcs of $N$, and $N_i$ be the sub-network of $N$ on $P_i$. The number of hybrid vertices of $N$ is equal to the sum of the number of hybrid vertices of each $N_i$ and the number of hybrid vertices of the simple network of $N$. So if $N$ is the network that minimizes the minimum number of hybrid vertices, then $N_i$ has to be also the network which minimizes the number of hybrid vertices among those who are consistent with $\mathcal{T}|P_i$. This property allows us to have a recursive construction as the algorithm \ref{algo}. Indeed, in the algorithm \ref{algo}, for every node $a$ of $R$, instead of taking any simple network consistent with $\mathcal{T}' \nabla \mathcal{P}$, we take the one such that the corresponding network minimizes the number of hybrid vertices. In the end, $N_r$, where $r$ is the root of $R$, will be the wanted network. The construction described in the algorithm \ref{algo_min} stays in polynomial time because we can find all simple level-k networks in $O(|T|^{k+1})$ time, and all possible partitions of  the leaf set in $O(n^{3k})$ time. Finally, the recursion on $O(n)$ non singleton SN-set makes the total complexity $O(|\mathcal{T}|^{k+1}n^{3k+1})$.
\end{pf}

\section{Conclusion and perpectives}
To any  set of triplets $S$ we can define its $treerank(S)$ as the minimum $k$ for which it exist a level-k network which represents $S$. This measures the distance from $S$ to a tree in term of number of hybrid nodes.
We prove here that for dense triplets, for any fixed $k$, checking if $treerank(S)\leq k$ can be done in polynomial time. Therefore this new parameter is analogous to treewidth for graphs and we conjecture that its computation is NP-hard for dense triplets or extremely dense triplets.
However, comparing with the complexity of the existing efficient algorithms for the cases $k = 0, 1, 2$, a better bound can be expected for level-k networks.   Another interesting question is under which conditions on the triplet set $\mathcal{T}$ there is only one network $N$ consistent with $\mathcal{T}$. We also would like  to know if the condition of density on the triplet set can be relaxed  so that there is still a polynomial algorithm to construct a consistent level-k network, if there any, with any $k$ fixed.

\section{Thanks}
We would like to thank Philippe Gambette for many useful references and comments and also for his practical website \textit{Who is Who in Phylogenetic Networks} which let us to know what already exists about phylogenetic networks.

\bibliographystyle{plain}
\bibliography{phylogenetic_network}

\end{document}